\documentclass[aps,prc,twocolumn,superscriptaddress,showpacs,floatfix,nofootinbib]{revtex4-1}
\usepackage{amsmath,graphicx}
\newcommand{\avgev}[1]{\left\langle{#1}\right\rangle}
\def\bra{\langle}
\def\ket{\rangle}

\begin{document}

\title{$v_4$, $v_5$, $v_6$, $v_7$: nonlinear hydrodynamic response
  versus LHC data}

\author{Li Yan}
\author{Jean-Yves Ollitrault}
\affiliation{
Institut de physique th\'eorique, Universit\'e Paris Saclay, CNRS, CEA, F-91191 Gif-sur-Yvette, France} 
\date{\today}

\begin{abstract}
Higher harmonics of anisotropic flow ($v_n$ with $n\ge 4$) in
heavy-ion collisions can be measured either with respect to their own
plane, or with respect to a plane constructed using lower-order
harmonics. We explain how such measurements are related to event-plane
correlations. 
We show that CMS data on $v_4$ and $v_6$ are compatible with ATLAS data 
on event-plane correlations. 
If one assumes that higher harmonics are the superposition of
non-linear and linear responses, then the linear and 
non-linear parts can be isolated under fairly general assumptions. 
By combining analyses of higher harmonics with analyses of $v_2$ and
$v_3$, one can eliminate the uncertainty from initial conditions and
define quantities that only involve nonlinear hydrodynamic response
coefficients.  
Experimental data on $v_4$, $v_5$ and $v_6$ are in good agreement with
hydrodynamic calculations. 
We argue that $v_7$ can be measured with respect to elliptic and
triangular flow. We present predictions for $v_7$ versus centrality in Pb-Pb
collisions at the LHC. 
\end{abstract}


\maketitle

\section{Introduction}
In the last year or so, LHC and RHIC experiments have probed anisotropic
flow~\cite{Heinz:2013th} 
and its fluctuations~\cite{Alver:2006wh,Alver:2010gr} 
to an unprecedented degree of 
precision~\cite{Chatrchyan:2013kba,Aad:2014fla,Abelev:2014mda,Aad:2014vba,Adare:2014kci}. 
These new analyses include in particular detailed analyses of higher
Fourier harmonics ($v_4$, $v_5$, $v_6$) and their correlations with
lower harmonics ($v_2$, $v_3$). 
The scope of this paper is twofold. The first goal is to point out
specific relations between seemingly different observables found in
the recent experimental literature, and to propose new observables. 
The second goal is to show that measurements of higher harmonics can
be combined with measurements of lower harmonics 
in a way that facilitates comparison with theory. 
As an illustration, recent experimental results are compared with
hydrodynamic calculations. 

The CMS Collaboration has measured $v_4$ and $v_6$ with respect to
their own direction, and with respect to the direction of elliptic
flow $v_2$~\cite{Chatrchyan:2013kba} (see also \cite{Adare:2014kci});
on the other hand, the   
ATLAS Collaboration has measured a large number of event-plane
correlations~\cite{Aad:2014fla}. 
In Sec.~\ref{s:differentvn}, we clarify the relation between these
observables and show how they are related to one another. In
particular, we show that CMS and ATLAS data on $v_4$ and $v_6$ are
compatible. We explain how odd harmonics, such as $v_5$ or $v_7$,
can also be analyzed with respect to the direction of lower
harmonics. 

While recent experimental data have been compared to several
theoretical models, either 
event-by-event hydrodynamic
calculations~\cite{Schenke:2011bn,Gardim:2012yp,Qiu:2012uy,Ryu:2015vwa} 
or tranport models~\cite{Bhalerao:2013ina}, these comparisons offer
little insight into the physics of higher-order harmonics. 
In particular, theoretical calculations depend strongly 
on the model of the initial density profile, which has long been
recognized as the main source of uncertainty in modeling anisotropic 
flow~\cite{Luzum:2008cw}. 
On the other hand, there are hints that the physics of higher-order
harmonics should be simple: for instance, 
the ratio $v_4/(v_2)^2$~\cite{Borghini:2005kd,Lang:2013oba} is 
equal to $\frac{1}{2}$ at high transverse momentum $p_T$ in ideal hydrodynamics. 

In hydrodynamics, higher-order harmonics are superpositions of linear and
non-linear response
terms~\cite{Gardim:2011xv,Teaney:2012ke,Teaney:2013dta,Gardim:2014tya}. 
This is recalled in Sec.~\ref{s:linearnonlinear}. 
We explain how the linear and nonlinear terms can be isolated 
under fairly general assumptions. 
We show how analyses of higher-order harmonics can be combined 
with analyses of lower-order harmonics ($v_2$ and $v_3$) to form
quantities which do not involve the initial state. These quantities 
are compared with hydrodynamic calculations. 

In Sec.~\ref{s:prediction}, we list a few predictions for
higher-order harmonics; in particular, we predict the value of $v_7$,
measured with respect to $v_2$ and $v_3$, as a function of centrality. 

\section{Observables for higher harmonics}
\label{s:differentvn}

Anisotropic flow is
an azimuthal ($\varphi$) asymmetry of the single-particle distribution~\cite{Luzum:2011mm}:
\begin{equation}
\label{defVn}
P(\varphi)=\frac{1}{2\pi}\sum_{n=-\infty}^{+\infty}V_n e^{-in\varphi},
\end{equation}
where $V_n=v_n\exp(in\Psi_n)$ is the (complex) anisotropic flow coefficient in the $n$th
harmonic, and $V_{-n}=V_n^*$. 
Both the magnitude~\cite{Miller:2003kd} and
phase~\cite{Andrade:2006yh,Alver:2006wh} of $V_n$ fluctuate event to event. 

The simplest observable involving $V_n$ is a plain rms
average~\cite{ALICE:2011ab,Adare:2011tg}:
\begin{equation}
\label{eq:vnpsin}
v_n\{\Psi_n\}\equiv\sqrt{\langle |V_n|^2\rangle},
\end{equation}
where angular brackets denote an average over events. 
The notation $v_n\{\Psi_n\}$ has been used earlier to denote 
the value analyzed with the event-plane
method~\cite{Chatrchyan:2013kba}. However, the event-plane method does
does not quite measure the rms average~\cite{Alver:2008zza}.
Therefore it should be replaced by the scalar-product
method~\cite{Luzum:2012da}, which is recalled in
Appendix~\ref{s:appendix}. 
Note that our $v_n\{\Psi_n\}$ is the same quantity as $v_n\{2\}$
in the notation of the cumulant analysis~\cite{Borghini:2001vi}. 

Alternatively, $V_4$ can be analyzed with respect to 
the direction of $V_2$~\cite{Adams:2003zg,Adare:2010ux}, and 
$V_6$ can be analyzed with respect to the direction of $V_2$ or that
of  $V_3$, 
 \begin{eqnarray}
\label{eq:v4psi2}
v_{4}\{\Psi_2\} &\equiv& \frac{{\rm Re}\bra V_4 (V_2^{*})^2\ket}{\sqrt{\bra|V_2|^4
  \ket}}\cr
v_{6}\{\Psi_2\} &\equiv& \frac{{\rm Re}\bra V_6 (V_2^{*})^3\ket}{\sqrt{\bra|V_2|^6
  \ket}}\cr
v_{6}\{\Psi_3\} &\equiv& \frac{{\rm Re}\bra V_6 (V_3^{*})^2\ket}{\sqrt{\bra|V_3|^4
  \ket}}.
\end{eqnarray}
The triangular inequality implies $|v_{4}\{\Psi_2\}|\le v_4\{\Psi_4\}$, 
$|v_{6}\{\Psi_2\}|\le v_6\{\Psi_6\}$, 
$|v_{6}\{\Psi_3\}|\le v_6\{\Psi_6\}$, 
i.e., $v_4$ and $v_6$ are larger
when measured with respect to their own plane than with respect to
another plane. 
The ratio of $v_{n}\{\Psi_{m}\}$ and $v_n\{\Psi_n\}$ (where $n$ is a
multiple of $m$)
can be written as the 
Pearson correlation coefficient between $V_n$ and $(V_m)^{n/m}$, which
we denote by $\rho_{mn}$: 
\begin{eqnarray}
\label{eq:pc24}
\rho_{24} &\equiv& \frac{{\rm Re}\bra V_4 (V_2^{*})^2\ket}{\sqrt{\bra|V_4|^2
  \ket\bra|V_2|^4\ket}}=\frac{v_{4}\{\Psi_2\}}{v_4\{\Psi_4\}}\cr
\rho_{26} &\equiv& \frac{{\rm Re}\bra V_6 (V_2^{*})^3\ket}{\sqrt{\bra|V_6|^2
  \ket\bra|V_2|^6
  \ket}}=\frac{v_{6}\{\Psi_2\}}{v_6\{\Psi_6\}}\cr
\rho_{36} &\equiv& \frac{{\rm Re}\bra V_6 (V_3^{*})^2\ket}{\sqrt{\bra|V_6|^2
  \ket\bra|V_3|^4
  \ket}}=\frac{v_{6}\{\Psi_3\}}{v_6\{\Psi_6\}}.
\end{eqnarray}
The correlations between event planes measured by ATLAS,
which are denoted by $\avgev{\cos(4(\Phi_2-\Phi_4))}_w$, 
$\avgev{\cos(6(\Phi_2-\Phi_6))}_w$ 
and $\avgev{\cos(6(\Phi_3-\Phi_6))}_w$ 
in Ref.~\cite{Aad:2014fla}, are {\it precisely\/} 
$\rho_{24}$, 
$\rho_{26}$ and $\rho_{36}$~\cite{Bhalerao:2013ina,Bhalerao:2014xra}. 
Note that the terminology ``event-plane correlations'' applied to such
measurements is somewhat misleading, in the sense that these observables
involve not only the angles of $V_n$, but also their
magnitudes~\cite{Luzum:2012da}.
 
\begin{figure}
\begin{center}
\includegraphics[width=0.8\linewidth]{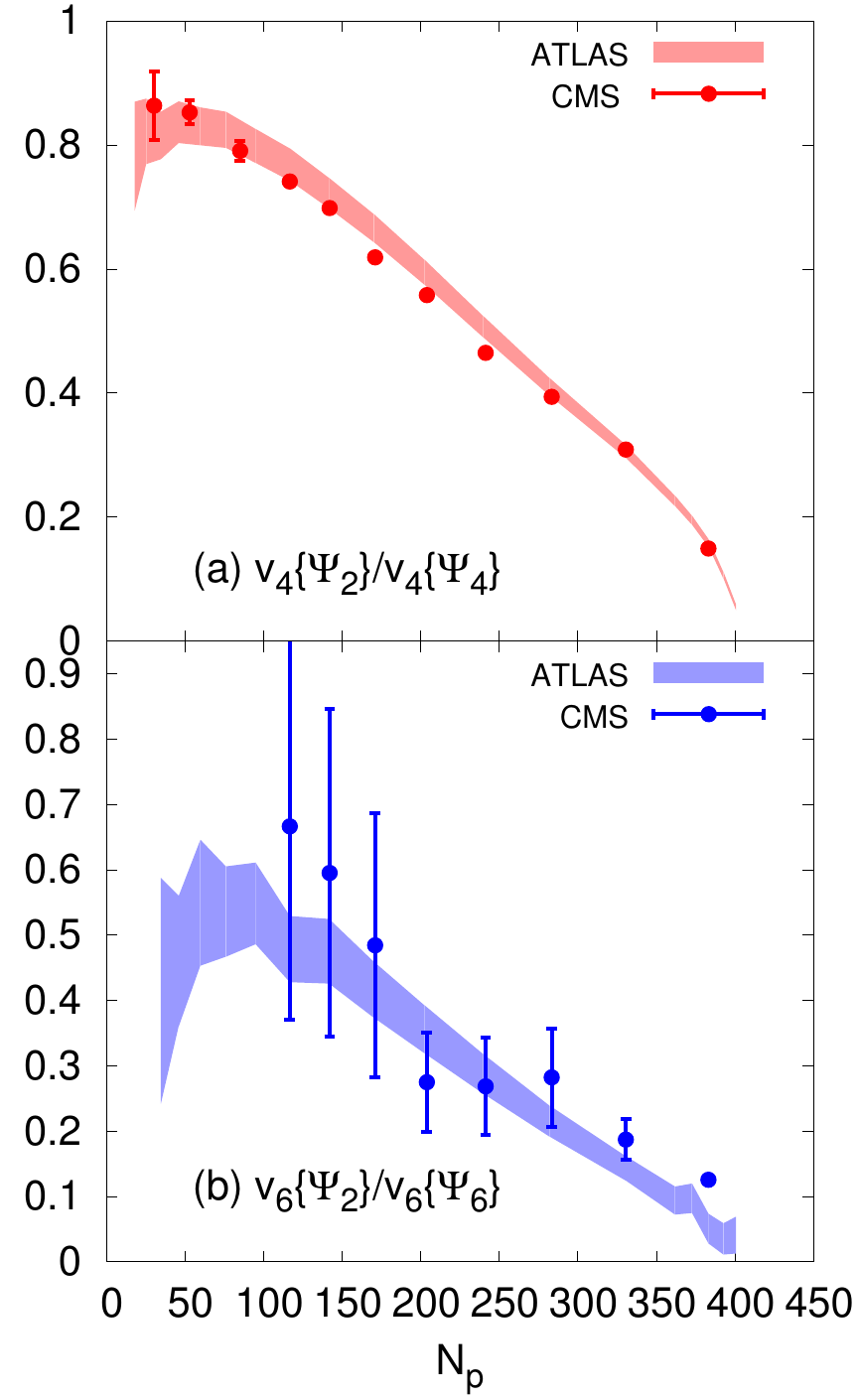}
\caption{(Color online) 
Test of Eqs.~(\ref{eq:pc24}). 
Shaded bands correspond to the left-hand side
measured by ATLAS~\cite{Aad:2014fla} in Pb-Pb collisions at $2.76$~TeV. 
Full circles correspond to the 
right-hand side, obtained using CMS data~\cite{Chatrchyan:2013kba}. }
\label{fig:fig1}
\end{center}
\end{figure}
Figure~\ref{fig:fig1} presents a test of the first two lines of
Eq.~(\ref{eq:pc24}),  
where the left-hand side uses ATLAS data and the right-hand side CMS
data.  
The overall agreement is very good, which shows that CMS and ATLAS
data are compatible,
even though they are measured with different cuts of tranverse momentum $p_T$.
Note that CMS uses the event-plane method, instead of the
scalar-product method. This method yields a slightly lower correlation
when the 
resolution is large~\cite{Aad:2014fla}. This explains, at least
qualitatively, why CMS data are slightly lower than ATLAS data for
midcentral collisions in Fig.~\ref{fig:fig1} (a).   

While Pearson correlation coefficients are typically analyzed by
integrating over all particles in a reference
detector~\cite{Aad:2014fla}, 
analyses of $v_n$ with respect to a specific direction (either
$\Psi_2$ or $\Psi_n$) can be done differentially, as a function of
transverse momentum $p_T$~\cite{Adare:2010ux} (see
Appendix~\ref{s:appendix} for analysis details).  
Hydrodynamics predicts a slightly different $p_T$ dependence 
depending on the reference direction~\cite{Teaney:2012ke}. 
It is therefore interesting to generalize Eq.~(\ref{eq:v4psi2}) to odd 
harmonics. $V_5$ and $V_7$ can be analyzed with respect to the
directions of $V_2$ and $V_3$ in the following way: 
 \begin{eqnarray}
\label{eq:v5psi23}
v_{5}\{\Psi_{23}\} &\equiv& \frac{{\rm Re}\bra V_5 V_2^*V_3^*\ket}{\sqrt{\bra|V_2|^2|V_3|^2
  \ket}}\cr
v_{7}\{\Psi_{23}\} &\equiv& \frac{{\rm Re}\bra V_7 (V_2^{*})^2V_3^*\ket}{\sqrt{\bra|V_2|^4|V_3|^2
  \ket}}.
\end{eqnarray}
Quantitative predictions for these quantities will be presented in
Sec.~\ref{s:prediction}. 
These projected harmonics are smaller than those defined by 
Eq.(\ref{eq:vnpsin}), namely, $|v_{5}\{\Psi_{23}\}|\le v_5\{\Psi_5\}$, 
(and $|v_{7}\{\Psi_{23}\}|\le v_7\{\Psi_7\}$). 
The ratio of $|v_{5}\{\Psi_{23}\}|$ and $v_5\{\Psi_5\}$ is again the
Pearson correlation coefficient between $V_5$ and $V_2V_3$:
\begin{equation}
\label{eq:pc235}
\rho_{23,5} \equiv\frac{{\rm Re} \bra V_5 V_2^*V_3^*\ket}{\sqrt{\bra |V_2|^2|V_3|^2 \ket\bra |V_5|^2\ket}}=\frac{v_{5}\{\Psi_{23}\}}{v_5\{\Psi_5\}}.
\end{equation}
This quantity is very similar to the corresponding three-plane
correlation measured by ATLAS~\cite{Aad:2014fla}:
\begin{equation}
\label{eq:atlas235}
\bra\cos(2\Phi_2+3\Phi_3-5\Phi_5)\ket_w
\equiv\frac{{\rm Re} \bra V_5 V_2^*V_3^*\ket}{\sqrt{\bra |V_2|^2\ket\bra|V_3|^2 \ket\bra |V_5|^2\ket}}.
\end{equation}
More precisely, they coincide if the magnitudes of $V_2$ and $V_3$ are
uncorrelated,\footnote{ 
A slight anticorrelation between $|V_2|^2$ and $|V_3|^2$ has been
predicted in AMPT simulations~\cite{Huo:2013qma,Bilandzic:2013kga,Bhalerao:2014xra}, but
it is at most at the 10\% level.}  namely, $\bra |V_2|^2|V_3|^2 \ket=\bra
|V_2|^2\ket\bra|V_3|^2 \ket$. 
Throughout this paper, we use $\bra\cos(2\Phi_2+3\Phi_3-5\Phi_5)\ket_w$ 
from ATLAS as an approximation for $\rho_{235}$. 

Note that even though $v_4\{\Psi_{2}\}$ and $v_6\{\Psi_{2}\}$ are
smaller than $v_4\{\Psi_{4}\}$ and $v_6\{\Psi_{6}\}$, respectively,
they are measured with better relative precision~\cite{Chatrchyan:2013kba}. 
The reason is that these measurements use elliptic flow as a reference,
which is measured very accurately. 
Triangular flow, $v_3$, is also precisely known. 
We therefore expect that $v_{5}\{\Psi_{23}\}$ 
be determined with better relative accuracy than 
$v_5\{\Psi_{5}\}$.  
In the same way, we expect that even though no experiment has yet been
able to detect a nonzero $v_{7}\{\Psi_{7}\}$, LHC experiments could 
already measure $v_{7}\{\Psi_{23}\}$. 

\section{Linear and nonlinear response}
\label{s:linearnonlinear}

In hydrodynamics, anisotropic flow is the response to anisotropy in
the initial density profile~\cite{Floerchinger:2013tya}. Harmonics
$V_4$ and higher can arise  
from initial anisotropies in the same
harmonic~\cite{Alver:2010gr,Teaney:2010vd,Gubser:2010ui,Hatta:2014jva} 
(linear response) or can be induced by lower-order 
harmonics~\cite{Borghini:2005kd,Bravina:2013xla,Bravina:2013ora}
(nonlinear response).  
To a good approximation~\cite{Gardim:2014tya}, one can write
\begin{eqnarray}
\label{decomposition}
V_4&=&V_{4L}+\chi_4 (V_2)^2\cr
V_5&=&V_{5L}+\chi_5 V_2V_3\cr
V_6&=&V_{6L}+\chi_{62} (V_2)^3+\chi_{63}(V_3)^2\cr
V_7&=&V_{7L}+\chi_7 (V_2)^2V_3,
\end{eqnarray} 
where $V_{nL}$ denotes the part of $V_n$ due to linear response, and
we have included the nonlinear terms involving the largest flow 
harmonics, $V_2$ and $V_3$. 
The interest of this decomposition is that the nonlinear response
coefficients $\chi$ are independent of the initial density profile in
a given centrality class~\cite{Teaney:2012ke}.   
We now explain how the linear and nonlinear parts can be isolated.

\subsection{Linear response}

The linear part of $v_4$ and $v_5$ 
can be isolated~\cite{Jia:2014jca} by combining the 
observables introduced in Sec.~\ref{s:differentvn}.
Using Eqs.~(\ref{eq:vnpsin}) and (\ref{eq:v4psi2}), one obtains
\begin{eqnarray}
\label{eq:projV4}
(v_4\{\Psi_4\})^2-(v_4\{\Psi_2\})^2&=&\bra|V_{4L}|^2\ket-
\frac{|\bra V_{4L}(V_2^*)^2\ket|^2}{\bra |V_2|^4\ket}\cr
(v_5\{\Psi_5\})^2-(v_5\{\Psi_{23}\})^2&=&\bra|V_{5L}|^2\ket-
\frac{|\bra V_{5L}V_2^*V_3^*\ket|^2}{\bra |V_2|^2|V_3|^2\ket}. 
\end{eqnarray}
These results are general: these combinations always subtract
the nonlinear response. 

From now on, we further assume that the terms appearing in the  
right-hand side of Eq.~(\ref{decomposition}) are uncorrelated. 
That is, we neglect the small correlation between the linear and
nonlinear parts which is seen in Monte-Carlo Glauber 
simulations~\cite{Teaney:2012ke}. 
The idea behind this assumption is that $V_{4L}$ is produced by 
initial fluctuations in the fourth harmonic, which are not correlated
with the mean eccentricity. 
Then, the last term in the right-hand side of Eq.~(\ref{eq:projV4})
vanishes, and the rms value of the linear part is 
\begin{eqnarray}
\label{eq:projV42}
v_{4L}&\equiv&\sqrt{\bra|V_{4L}|^2\ket}=\sqrt{(v_4\{\Psi_4\})^2-(v_4\{\Psi_2\})^2}\cr
v_{5L}&\equiv&\sqrt{\bra|V_{5L}|^2\ket}=\sqrt{(v_5\{\Psi_5\})^2-(v_5\{\Psi_{23}\})^2}.
\end{eqnarray}
This quantity has been measured as a function of centrality by the 
ATLAS collaboration~\cite{Jia:2014jca}. 

\subsection{Nonlinear response}

\begin{figure*}
\includegraphics[width=\linewidth]{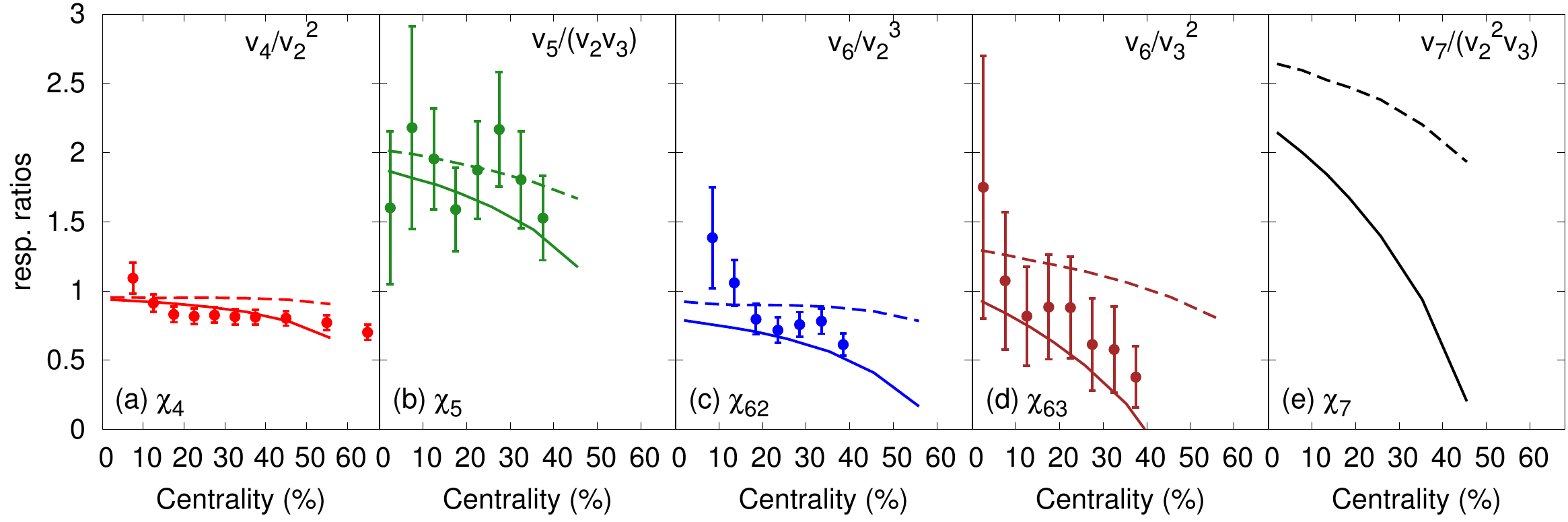}
\caption{ (Color online) 
Nonlinear response coefficients defined by Eq.~(\ref{nonlinear}) as a
function of centrality. Each panel corresponds to a different line of
Eq.~(\ref{nonlinear}). 
Dashed lines: ideal hydrodynamics. 
Solid lines: viscous hydrodynamics with $\eta/s=0.08$. 
Symbols: experimental data (see text for details). }
\label{fig:resp-ratios}
\end{figure*}
The nonlinear parts are obtained by projecting
Eq.~(\ref{decomposition}) onto lower harmonics. 
Assuming again that the terms in the right-hand side of
Eq.~(\ref{decomposition}) are uncorrelated, one obtains the following
expressions for nonlinear response coefficients:
\begin{eqnarray}
\label{nonlinear}
\chi_4&=&\frac{\bra V_4 (V_2^*)^2\ket}{\bra |V_2|^4\ket}
=\frac{v_4\{\Psi_2\}}{\sqrt{\bra |V_2|^4\ket}}
\cr
\chi_5&=&\frac{\bra V_5 V_2^*V_3^*\ket}{\bra |V_2|^2 |V_3^2|\ket}
=\frac{v_5\{\Psi_{23}\}}{\sqrt{\bra |V_2|^2 |V_3^2|\ket}}\cr
\chi_{62}&=&\frac{\bra V_6 (V_2^*)^3\ket}{\bra |V_2|^6\ket}
=\frac{v_6\{\Psi_2\}}{\sqrt{\bra |V_2|^6\ket}}\cr
\chi_{63}&=&\frac{\bra V_6 (V_3^*)^2\ket}{\bra |V_3|^4\ket}
=\frac{v_6\{\Psi_3\}}{\sqrt{\bra |V_3|^4\ket}}\cr
\chi_7&=&\frac{\bra V_7(V_2^*)^2V_3^*\ket}{\bra |V_2|^4 |V_3^2|\ket}
=\frac{v_7\{\Psi_{23}\}}{\sqrt{\bra |V_2|^4 |V_3^2|\ket}}.
\end{eqnarray} 
The left-hand side of these expressions can be calculated in
hydrodynamics, and is independent of the model of initial conditions, 
while the right-hand side can be inferred from experimental data. 
Eq.~(\ref{nonlinear}) therefore offers a direct comparison between
hydrodynamics and data, where all dependence on initial
state is eliminated~\cite{Gombeaud:2009ye}. 
Nonlinear response coefficients have been obtained using
event-shape engineering~\cite{Jia:2014jca} by the ATLAS collaboration. 
The present method does not require event-shape engineering. 
The comparison between hydrodynamics and data is shown in
Fig.~\ref{fig:resp-ratios}, and we  
now explain in detail how these results are obtained. 

The numerators in the right-hand side of Eq.~(\ref{nonlinear}) are 
the projected harmonics defined by Eqs.~(\ref{eq:v4psi2}) and
(\ref{eq:v5psi23}). 
We use $v_4\{\Psi_2\}$ and  $v_6\{\Psi_2\}$ measured by 
CMS~\cite{Chatrchyan:2013kba}.\footnote{Since CMS uses the 
  event-plane method, the results are slightly lower than the nominal
  quantities in Eqs.~(\ref{eq:v4psi2})  \cite{Luzum:2012da}.} 
$v_5\{\Psi_{23}\}$ and $v_6\{\Psi_3\}$ are not measured directly, but
can be inferred from $v_5\{\Psi_{5}\}$, $v_6\{\Psi_6\}$, 
$\rho_{235}$ and $\rho_{36}$ using Eqs.~(\ref{eq:pc24}) and
(\ref{eq:pc235}). 
We use CMS data~\cite{Chatrchyan:2013kba} for $v_5\{\Psi_{5}\}$ and
$v_6\{\Psi_6\}$ and ATLAS data~\cite{Aad:2014fla} for $\rho_{235}$ and
$\rho_{36}$. These correlation coefficients, however, are expected to depend
little on the experimental setup, as illustrated in 
Fig.~\ref{fig:fig1}.

The denominators in the right-hand side of Eq.~(\ref{nonlinear})
involve various even moments of the distribution of 
$V_2$ and $V_3$. There is no direct measurement of these moments to
date. A straightforward procedure to analyze them is outlined in
Ref.~\cite{Bhalerao:2014xra}.  
Alternatively, 
moments of the form $\bra |V_n|^{2k}\ket$ can be inferred from
cumulants~\cite{Borghini:2001vi}.  
The expressions of the first moments in terms of cumulants are: 
\begin{eqnarray}
\label{moments}
\bra|V_n|^2\ket &=& v_2\{2\}^2\cr
\bra|V_n|^4\ket &=& 2v_2\{2\}^4-v_2\{4\}^4\cr
\bra |V_n|^6\ket &=& 4v_n\{6\}^6 - 9v_n\{4\}^4v_n\{2\}^2+6v_n\{2\}^6.
\end{eqnarray}   
For the moments involving both $V_2$ and $V_3$ (second and fourth line of
Eq.~(\ref{nonlinear})), 
we further assume that the magnitudes of $V_2$ and $V_3$ are
uncorrelated. 

Since different experiments have different acceptance (in particular
in transverse momentum $p_T$), it is important to use results from the
same experiment in evaluating the right-hand side of
Eq.~(\ref{nonlinear}). 
We use cumulant results from CMS~\cite{Chatrchyan:2013kba}. 
CMS has not published $v_2\{6\}$, but ATLAS has observed~\cite{Aad:2014vba}
that $v_2\{6\}\simeq v_2\{4\}$ for all centralities, therefore we
assume $v_2\{6\}=v_2\{4\}$. 

The response coefficients in the left-hand side of
Eq.~(\ref{nonlinear}) are calculated using hydrodynamics. 
The calculation shown in Fig.~\ref{fig:resp-ratios} is the same as 
in Ref.~\cite{Teaney:2012ke}. 
It uses as initial condition a symmetric Gaussian density profile, 
where the normalization is adjusted to fit the measured multiplicity
$dN_{ch}/dy$ of Pb-Pb collisions at the LHC in the corresponding
centrality class. 
This symmetric profile 
is deformed in order to produce anisotropic flow in the desired
harmonic.\footnote{For instance, $\chi_4$ is obtained by introducing an
elliptic deformation and calculating $\chi_4=v_4/(v_2)^2$.} 
We assume uniform longitudinal expansion~\cite{Bjorken:1982qr}. 
With these initial conditions, we solve ideal hydrodynamics or second
order viscous hydrodynamics~\cite{Baier:2007ix} with 
constant shear viscosity over entropy ratio
$\eta/s=0.08$~\cite{Kovtun:2004de}. 
The equation of state is taken from Lattice QCD~\cite{Laine:2006cp}. 
The initial time of the calculation is $\tau_o=1$~fm/$c$  
and the freeze-out temperature~\cite{Kolb:2003dz} is
$T_{fo}=150$~MeV. 
Anisotropic flow, $v_n$, is calculated at freeze-out. 
It is averaged over particles in the interval 
$p_T>0.3$~GeV/$c$, corresponding to the CMS 
acceptance~\cite{Chatrchyan:2013kba}. 

Figure~\ref{fig:resp-ratios} shows that hydrodynamics
naturally captures the sign, the magnitude, and the centrality
dependence of all four nonlinear response coefficients. 
Experimental results differ from hydrodynamic calculations only for
the most central bins~\cite{Gombeaud:2009ye}, where the linear part
typically becomes larger than the nonlinear part and their correlation
can no longer be neglected.

The order of magnitude of the hydrodynamic result can be 
understood simply. 
At fixed, large $p_T$, ideal hydrodynamics 
predicts~\cite{Borghini:2005kd,Teaney:2012ke}  
$\chi_4=\frac{1}{2}$, 
$\chi_5=1$, 
$\chi_{62}=\frac{1}{6}$, 
$\chi_{63}=\frac{1}{2}$, 
$\chi_{7}=\frac{1}{2}$.
However, after averaging over $p_T$,  $\chi_4$ is multiplied by 
$\langle v_2^2\rangle/\langle v_2\rangle^2>1$, where brackets now
denote an average over $p_T$ in a single hydro event. This is the reason why
the results  
shown in Fig.~\ref{fig:resp-ratios} are larger than the fixed-$p_T$ 
prediction. 
Since $v_2$ and $v_3$ have similar $p_T$ dependences, 
the enhancement factor is roughly the same 
for all quadratic response  
terms: panels (a), (b), (d) show that $\chi_{5}\sim 2\chi_4$,
$\chi_{63}\sim \chi_4$,  in agreement with the above values. 
The enhancement from averaging over $p_T$ is larger for cubic response
terms than for quadratic terms, but it is
similar for both cubic terms:
panels (c) and (e) show that $\chi_{7}\sim 3\chi_{62}$, also in agreement
with the above values. 

A full hydrodynamical calculation gives results which differ somewhat
from the naive predictions above.
Coefficients from ideal hydrodynamics have a slight centrality
dependence which is not captured by these formulas~\cite{Luzum:2010ae}. 
Viscous hydrodynamics predicts lower coefficients than ideal
hydrodynamics. 
Effect of viscosity, however, cancel to a large extent in the ratios:
they are much smaller on $\chi_4=v_4/(v_2)^2$ than on $v_4$ and
$(v_2)^2$ individually~\cite{Teaney:2012ke}.  

Nonlinear response coefficients are mostly determined at 
freeze-out~\cite{Teaney:2012ke}, which is probably the least
understood part of hydrodynamic calculations. 
While they depend little on the details of the
initial profile or of the hydrodynamic evolution, they depend rather
strongly on the freeze-out temperature~\cite{Luzum:2010ae}. Similarly,
the dependence of our results on viscosity is mostly through the
viscous correction 
to the momentum distribution at freeze-out~\cite{Teaney:2003kp}.  
The momentum distribution at freeze-out is not constrained
theoretically~\cite{Dusling:2009df,Bhalerao:2013pza}, and the
quadratic ansatz used in this calculation is not favored by previous studies of
$v_4$~\cite{Luzum:2010ad}. 
Our calculation does not involve bulk viscosity, which is likely to be
important at
freeze-out~\cite{Monnai:2009ad,Bozek:2009dw,Dusling:2011fd,Noronha-Hostler:2013gga}.  
Finally, our results are quite sensitive to the value of the
freeze-out temperature. 
Further studies are needed in order to pin down the sensitivity of
response coefficients to model parameters. 

\section{Predictions}
\label{s:prediction}

\begin{figure}
\begin{center}
\includegraphics[width=0.8\linewidth]{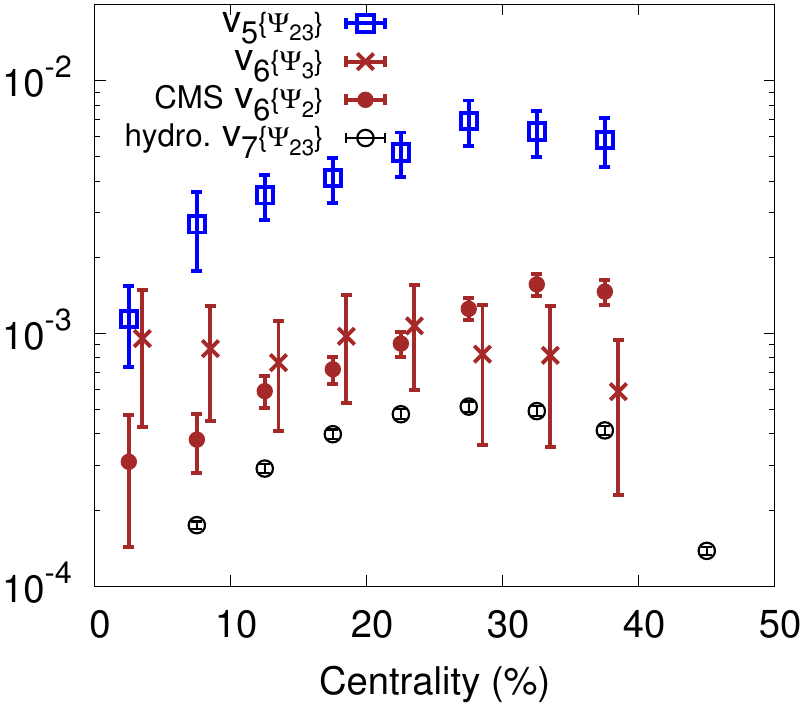}
\caption{(Color online) 
$v_5\{\Psi_{23}\}$, $v_6\{\Psi_{2}\}$, $v_6\{\Psi_3\}$ and
  $v_7\{\Psi_{23}\}$, averaged over charged particles with
  $p_T>0.3$~GeV/$c$, as a function of centrality in Pb-Pb collisions
  at $2.76$~TeV. $v_6\{\Psi_3\}$ has been shifted to the right by 1\%
  for sake of clarity.}
\label{fig:prediction}
\end{center}
\end{figure}
Figure~\ref{fig:prediction} displays $v_5\{\Psi_{23}\}$,
$v_6\{\Psi_{2}\}$, $v_6\{\Psi_3\}$ and $v_7\{\Psi_{23}\}$. 
Out of these four quantities, only $v_6\{\Psi_{2}\}$ has been measured by
CMS~\cite{Chatrchyan:2013kba}. 
We predict $v_5\{\Psi_{23}\}$ and $v_6\{\Psi_3\}$ using
Eqs.~(\ref{eq:pc24}) and (\ref{eq:pc235}), where we take 
$v_5\{\Psi_5\}$ and $v_6\{\Psi_6\}$ from CMS~\cite{Chatrchyan:2013kba}
and $\rho_{235}$ and $\rho_{36}$ from ATLAS~\cite{Aad:2014fla}. 

Finally, $v_7\{\Psi_{23}\}$ is obtained from the last line of
Eq.~(\ref{nonlinear}). We use the viscous hydrodynamic calculation
for $\chi_7$ shown in Fig.~\ref{fig:resp-ratios} (e). 
We again assume that the magnitudes of $V_2$ and $V_3$ are independent, 
that is,  $\bra |V_2|^4 |V_3^2|\ket\simeq \bra |V_2|^4\ket\bra
|V_3^2|\ket$, and we estimate the moments using Eq.~(\ref{moments})
and CMS data~\cite{Chatrchyan:2013kba}. 
We anticipate that the absolute experimental error on 
$v_7\{\Psi_{23}\}$ should be similar to the error on
$v_6\{\Psi_{2}\}$. 
This error is of the order of $0.01$\%. The predicted values of 
$v_7\{\Psi_{23}\}$ is $0.05$\% in the 25-30\% centrality range, 
larger than the error. We therefore expect that a nontrivial
$v_7\{\Psi_{23}\}$ could be measured in midcentral Pb-Pb collisions
at the LHC.  

\section{Conclusion}

Harmonics $v_4$ and higher can be measured either with respect to
their own planes or with respect to lower harmonic planes. 
We have clarified the relation between these projected harmonics and the
so-called  event-plane correlations. 

We have shown that under fairly general assumptions, measurements of
higher harmonics can be combined with measurements of $v_2$ and $v_3$
in a way that eliminates the dependence on the initial state, and can
be directly compared with hydrodynamic calculations. 
Experimental results for $v_4$, $v_5$ and $v_6$ are in good
agreement with viscous hydrodynamic calculations. 
We have argued that
$v_7$ could be measured, and presented quantitative predictions. 

On the experimental side, analyses should be repeated using 
the scalar-product method, whose result may differ significantly from
the event-plane method for higher harmonics~\cite{Luzum:2012da}. 
On the theoretical side, we hope that studies of higher harmonics will
help constrain the theoretical description of the fluid close the
freeze-out temperature, which is poorly understood at present. 

\begin{acknowledgments}
LY is funded  by the European Research Council under the 
Advanced Investigator Grant ERC-AD-267258. 
\end{acknowledgments}

\appendix
\section{Analysis}
\label{s:appendix}

The flow observables in Eq.~(\ref{eq:vnpsin}), (\ref{eq:v4psi2}) and
(\ref{eq:v5psi23}) are expressed in terms of moments of the
distribution of $V_n$. 
A generic moment is of the form~\cite{Bhalerao:2014xra}
\begin{equation}
\label{defmoments}
{\cal M}\equiv
\left\langle \prod_n {(V_n)^{k_n} (V_n^*)^{l_n}}\right\rangle,
\end{equation}
where $k_n$ and $l_n$ are integers and 
azimuthal symmetry implies 
$\sum_n n k_n=\sum_n n l_n$.
For instance, $\bra V_4(V_2^*)^2\ket$ corresponds to $k_4=1$,
$l_2=2$; $\bra |V_2|^6\ket$ corresponds to $k_2=l_2=3$; 
$\bra |V_2|^4|V_3|^2\ket$ corresponds to $k_2=l_2=2$,  $k_3=l_3=1$. 

We now describe a simple procedure for measuring these
moments~\cite{Bhalerao:2014xra}, which generalizes the 
scalar-product method~\cite{Adler:2002pu}. 
We define in each collision the flow vector by
\begin{equation}
\label{defqcomplex}
Q_n 
\equiv\frac{1}{N} \sum_j e^{in\varphi_j},
\end{equation} 
where the sum runs over $N$
particles seen in a reference detector, and $\varphi_j$ are their
azimuthal angles.
One measures $Q_n$ in two different parts of the detector 
(``subevents'') $A$ and $B$, 
which are symmetric around midrapidity and separated by a gap in 
pseudorapidity in order to suppress nonflow 
correlations~\cite{Adler:2003kt,Luzum:2010sp,Bhalerao:2013ina}. 
The moment (\ref{defmoments}) is then given by 
\begin{equation}
\label{qmoments}
{\cal M}=\left\langle \prod_n {(Q_{nA})^{k_n} (Q_{nB}^*)^{l_n}}\right\rangle.
\end{equation}
Applied to  Eq.~(\ref{eq:v5psi23}), this gives:
\begin{equation}
\label{eq:v5psi23int}
v_{5}\{\Psi_{23}\} \equiv \frac{{\rm Re}\bra Q_{5A}
  Q_{2B}^{*}Q_{3B}^{*}\ket}{\sqrt{{\rm Re}\bra  Q_{2A}Q_{3A}Q_{2B}^{*}Q_{3B}^{*}  \ket}}. 
\end{equation}
The scalar-product method thus uses the magnitude of the flow
vector~\cite{Adler:2002pu} while the traditional event-plane 
method~\cite{Poskanzer:1998yz} only uses its azimuthal angle. 
One can symmetrize the numerator of Eq.~(\ref{eq:v5psi23int}) over $A$
and $B$ to decrease the statistical error.  
Instead of 2 symmetric subevents, one can use 3 non-symmetric
subevents, as described in Ref.~\cite{Luzum:2012da}. 

Finally, analyses can be done differentially (in $p_T$ bins, for
identified particles, etc.). 
For the differential analysis, one replaces Eq.~(\ref{eq:v5psi23int}) by: 
\begin{equation}
\label{eq:v5psi3diff}
v_{5}\{\Psi_{23}\} \equiv \frac{{\rm Re}\bra e^{5i\varphi}
  Q_{2B}^{*}Q_{3B}^{*}\ket}{\sqrt{{\rm Re}\bra  Q_{2A}Q_{3A}Q_{2B}^{*}Q_{3B}^{*}  \ket}},
\end{equation}
where the average in the numerator is now an average over particles in
the considered bin, with azimuthal angle $\varphi$,  
instead of an average over events.

\end{document}